# Deriving Weeklong Activity-Travel Dairy from Google Location History: Survey Tool Development and A Field Test in Toronto

**Melvyn Li**, MASc. Candidate
Department of Civil & Mineral Engineering
University of Toronto
35 St. George St, Toronto, ON M5S 1A4
Email: melvyn.li@utoronto.ca

**Kaili Wang**, Ph.D. Student
Department of Civil & Mineral Engineering
University of Toronto
35 St. George St, Toronto, ON M5S 1A4
Email: jackkaili.wang@mail.utoronto.ca

**Yicong Liu**, Ph.D. Candidate
Department of Civil & Mineral Engineering
University of Toronto, Toronto, ON, Canada
Email: nora.liu@mail.utoronto.ca

**Khandker Nurul Habib**, Ph.D., P.Eng
Professor, Department of Civil & Mineral Engineering
University of Toronto
35 St. George St, Toronto, ON M5S 1A4
Email: khandker.nurulhabib@utoronto.ca

# ABSTRACT

This paper introduces an innovative travel survey methodology that utilizes Google Location History (GLH) data to generate travel diaries for transportation demand analysis. By leveraging the accuracy and omnipresence among smartphone users of GLH, the proposed methodology avoids the need for proprietary GPS tracking applications to collect smartphone-based GPS data. This research enhanced an existing travel survey designer, TRavel Activity Internet Survey Interface (TRAISI), to make it capable of deriving an activity-travel diary from respondent's GLH. The feasibility of this data collection approach is showcased through the Google Timeline Travel Survey (GTTS) conducted in the Greater Toronto Area, Canada. The resultant dataset from the GTTS is demographically representative and offers detailed and accurate travel behavioural insights.

**Keywords:** GPS survey; smartphone-based travel survey; Google Location History

# 1.0 Introduction

Transportation science is built upon analyzing people's everyday travel choices and behaviour. A key tool to understand travel patterns is the travel diaries which are sequential records of an individual's trips and activities over a specified period. Diaries collect valuable properties such as trip lengths, location names and addresses, and modes of transport. It provides the backbone for essential data in the study of travel behaviours.

In recent years, there have been attempts to utilize global positioning system (GPS) technology to provide a more efficient method of collecting travel diary information (Chen et al., 2010; Kelly et al., 2013; Shen & Stopher, 2014). Initially, dedicated GPS devices enabled the collection of accurate geolocation traces, including location, position, and speed parameters. This raw data must be professionally processed to obtain dataset useful for transportation analysis (Kerr et al., 2011; Paz-Soldan et al., 2014; Stopher et al., 2008). Moreover, data collection with dedicated GPS devices came with issues such as forgetting to bring or recharge the unit, as well as noise and loss of signal (Shen & Stopher, 2014; Kerr et al., 2011; Krenn et al., 2011; Paz-Soldan et al., 2014). In a comprehensive assessment of 24 research studies, Krenn et al. (2011) found that 17 of which reported GPS data loss attributed to a range of GPS devices and handling issues.

Nowadays, collection of GPS data has become significantly easier as a large portion of individuals now carry GPS capable smart devices in their daily lives (Stopher et al. 2007; Patterson & Fitzsimmons, 2016; Patterson et al., 2019). Along with improved cellular and GPS connection technology, issues with incomplete data may be minimized through the passive collection of location data using smartphones (Patterson & Fitzsimmons, 2016). The collected data contains more detail than traditional travel diaries and may be collected for longer periods of time (Allström et al., 2017; Molloy et al., 2022). The burden on each respondent is also greatly reduced when compared to traditional travel diary methods (Itsubo & Hato, 2006).

One barrier to conducting smartphone-based GPS surveys is developing, maintaining, and accessing smartphone applications that collect respondents' GPS traces. Some researchers have developed the application in-house (Patterson et al., 2019; Cottrill et al., 2013; Prelipcean et al., 2018). Others rely on commercial applications such as the rMove and paid for software access on demand (Calastri et al., 2020; Resource Systems Group, 2017). Both methods necessitate a significant financial commitment, which continues to be an obstacle for transportation researchers worldwide attempting to collect and use GPS data for travel demand analysis purposes.

This study presents a free-of-charge approach to collect smartphone-based GPS data using Google Location History (GLH). The Google Maps is a widely used navigation software with more than one billion users worldwide (Business of Apps, 2023). A large existing user base greatly increases the amount of travel information that may be obtained. This study demonstrates that the GLH data can also be converted into travel diaries for travel demand modelling purposes. This approach has the application potential to substantially lower the worldwide barrier to collecting and using GPS data for travel demand analysis research. This study enhances the TRAISI tool to extract travel diary from the GLH (Wang et al., 2023). It also reports a proof-of-concept survey that collected weeklong travel diaries of 290 individuals in the Greater Toronto Area (GTA), Canada. Statistical techniques are also used to examine factors that influence the accuracy of the travel mode inference algorithm GLH uses to annotate each trip leg.

The paper is organized as follows. **Section 2** provides literature review on the development of travel survey methods collecting travel diary. **Section 3** presents the proposed method of collecting travel diary using GLH. **Section 4** presents descriptive analysis and results of the proof-of-concept survey. **Section 5** concludes the study by summarising key findings and recommendations for further research.

**2.0 Literature Review**

At early stage, travel diaries were collected using forms sent in the mail or through telephone surveys (Axhausen et al., 2002; Stopher, 1992). These traditional methods might be inaccurate as they rely solely on the respondent to self-report their travel activities. The drawbacks are compounded rapidly if the period to be recorded in the diary becomes greater. Due to the length of these processes, individuals may have to be compensated for their time for the survey to be attractive. This has always led to a trade-off between response rate and survey costs. Furthermore, fielding a paper-based survey incurs additional material and labor costs (Prelipcean et al., 2018). As well, with the decline of landlines and the popularity of mobile phones, many households do not have a central household phone number which weakens the reach of telephone-based surveys (Strauts, 2010).

With the advent of electronic data collection methods, various attempts aimed to improve the reach, completeness, and accuracy of travel diaries. Surveys were conducted using computer assisted telephone interviews (CATI), computer assisted personal interviews (CAPI), and computer-assisted self-interviews (CASI) (Wolf, 2006). Cost savings may also be found by making use of electronic and web survey forms instead of physical paper ones (Prelipcean et al., 2018). However, these measures do not eliminate drawback of traditional travel diary collection methods as the underlying survey procedures are similar. One such structural issue is that trips are recalled by survey participants after they have occurred. Traditional travel diaries where the respondent submits information based on their memory are prone to errors such as missing or incorrect information as well as forgetting to declare certain trips outright (Scully et al., 2017).

On the other hand, Wolf (2006) demonstrated that the percentage of missed trips in traditional travel diaries ranged from 11% to 81% based on six household travel surveys utilizing CATI and GPS carried out in the United States between 2001 and 2004. In a different study by Stopher et al. (2007), 50 selected households undertaking the Sydney Household Travel Survey were recruited to take a simultaneous GPS survey. After the collected GPS information was compared to the travel diary results, it was concluded that travel demand was under-reported by about 7% in the travel diary (Stopher et al., 2007). Thus, these problems have pushed researchers to explore new methods for collecting travel data to mitigate these issues.

GPS technology is expected to act as a more efficient means of collecting travel diary data. Initially, dedicated GPS-based devices were distributed to survey participants to collect geolocation traces (Chen et al., 2010; Kelly et al., 2013; Shen & Stopher, 2014). With the prevalence of smart devices, transportation researchers started to use proprietary smartphone-based application to collect location traces (Stopher et al., 2007; Patterson & Fitzsimmons, 2016; Patterson et al., 2019; Allström et al., 2017; Prelipcean et al., 2018; Molloy et al., 2022). Recently, Cools et al. (2021) noticed that Google Location History (GLH) contained information including geolocation traces, time stamps for each activity and inferred travel mode, which are sufficient to construct travel diaries. Cools et al. (2021) collected five sets of GLH to examine how well the GLH captured pre-determined activity traces. They found that GLH would forgo locations that dwelled less than the mean of 355 seconds. This finding suggests that GLH captured primary daily activity events and forgo noisy events less than five minutes in the daily activity-travel schedule. Their finding suggests the suitability to use GLH data in the activity-travel scheduling analysis, which is the key usage for travel diaries. This study presents an approach to demonstrate the applicability to collect GLH using random sampling and reports the quality of data for travel behaviour analysis purposes.

**3.0 Collecting Google Location History (GLH) Using TRAISI**

The most important consideration in travel survey design is the survey process respondents will follow. Traditionally, GPS surveys were performed in two phases, an initial recruit phase for collecting demographic data and a later travel diary collection phase where actual trips were traced (Lynch et al., 2019). It was decided that the collection of GLH data would also be a two-step survey process. **Figure 1** presents the flow of the data collection process. The first phase of the survey process ensures that respondents have correctly set up Google Maps location services to collect their location history. Socioeconomic and household information will also be collected during this first phase for geolocation and sample representation verification. Then, respondents are expected to conduct their daily activity and travel carrying their GLH enabled smartphone during the data collection period.

After the designated data collection period, respondents will be invited to the second survey phase. Respondents will be asked to submit their GLH data in phase two. This will require respondents to download the GLH data from their Google account. The location history is exported onto their computer as the standard Keyhole Markup Language (KML) geographic files. Once KML files are exported, the respondents will be in full procession of the data. Then, they will submit the files to the survey team with full consent. This procedure will ensure the data transaction is conducted exclusively between the respondents who have the ownership of their own GLH data and the survey team.

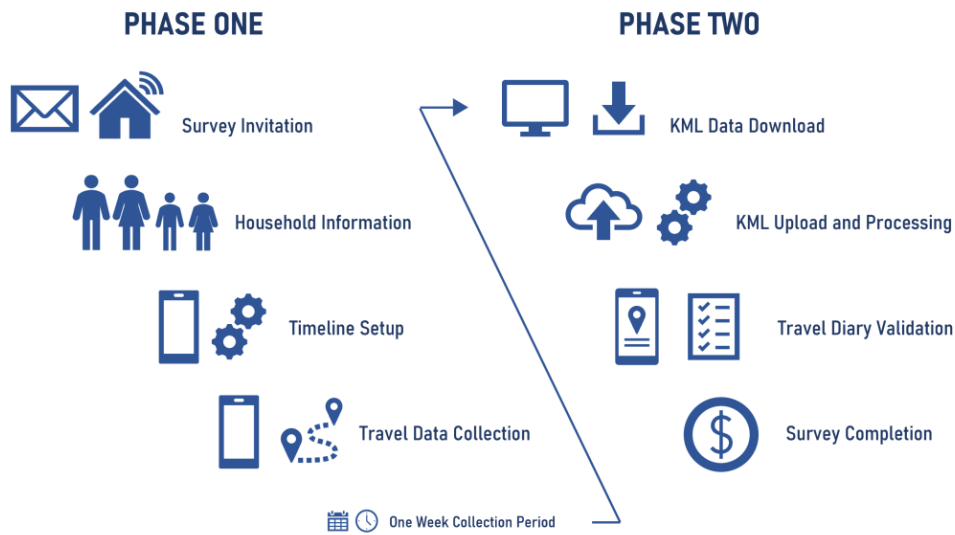

Figure 1. Schematic flow of Google Timeline Travel Survey (GTTS) process

The Travel and Activity Internet Survey Interface (TRAISI) survey platform is used to collect respondents' KML files (Wang et al., 2023). Once the GLH file is uploaded using the KML format, a completed travel diary will be generated which includes coordinate sequences, start and end times, durations of each event, names and locations of each activity, and inferred modes of each trip. Then users will then validate each activity purpose and trip travel mode in TRAISI. **Figure 2** presents an example of the automatically generated travel diary and the validation page in TRAISI. Respondents will repeat this process for every uploaded KML file forming comprehensive travel diaries for the study period.

Figure 2. Sample travel diary generated from Google Location History KML files in the TRAISI

**4.0 Google Timeline Travel Survey (GTTS)**

The survey procedure described above was utilized in a pilot study conducted in the Spring of 2023, between May and June, in the Greater Toronto Area (GTA), Canada. The study was named the Google Timeline Travel Survey (GTTS). The survey was randomly sampled from survey panels maintained by a commercial survey company. This section will discuss preliminary results of the survey. The distribution of socioeconomic variables in the GTTS will be compared with the 2021 Census (Statistics Canada, 2023). Moreover, trip and activity patterns from the GTTS dataset will be reviewed. Lastly, an explanatory analysis of the inferred travel mode in the GLH will be conducted by comparing the results against actual mode choice validated by respondents in the GTTS.

**Table 1** presents the descriptive statistics of the GTTS. In total, 57,531 individuals were invited to participate in the study. Of this group, 879 (1.53%) of the initial invitation successfully completed the first phase by sharing socioeconomic parameters and enabling their Google timeline. Moving on to the second phase, 290 individuals finished the entire survey process by accumulating one week of travel information, uploading their KML data files, and filling out the travel diaries. This amount is 0.50% of the initial invitation or 33.0% of the first phase respondents.

The representativeness of the GTTS is examined by comparing with the 2021 Census. Overall, the GTTS matched reasonably well with the population in the GTA described by the Census. As for gender distribution, more respondents were female, with 56.6% female versus 42.8% male. This trend may again be seen in the Census; however, the gap is larger in the GTTS dataset as the Census reported 51.2% female and 48.8% male.

The age distribution matched reasonably well between the GTTS and the Census. The GTTS only allowed individuals aged at least 18 years old to participate and only captured one individual under 20 years old. So, the age cohort comparison only focused on individuals older than 20 years old. The age distribution of the GTTS dataset was centered around the 30 to 39 age group, with a secondary peak in the 50 to 64 group. The mean age in the GTTS was 43.5 years old. The Census captured a greater number of senior individuals more than 65 years old, compared to the GTTS. In terms of household size, there were fewer single person households among the GTTS respondents compared to the Census; 13.1% in the GTTS compared to 25.0% in the Census.

For employment status, the majority of GTTS respondents (57.2%) were employed full-time. On the other hand, the Census captured much larger percentages of part-time employment and unemployment compared to the GTTS. This is consistent with the observation of age distribution. The Census captured more senior individuals aged 65 years old or older. On the other hand, the GTTS captured more working-age individuals.

Moving on to workplace arrangements, nearly half of the workers reported they could work remotely in the GTTS. The household income in the GTTS matched closely with the Census. The only discrepancy comes from households earning $200,000 and more. The GTTS underrepresented wealthy households which are less likely to participate in commercial survey panels.

Table 1. Overview of GTTS descriptive statistics

|  | Survey Stage | | |
| --- | --- | --- | --- |
|  | **Initial Invitation** | **Phase 1 - Setup** | **Phase 2 - Upload** |
| Number of respondents | 57,531 | 879 | 290 |
| % of Initial Invitation | - | 1.53% | 0.50% |
| % of Phase 1 Respondents | - | - | 33.0% |

|  | | **2023 GTTS** | **2021 Census** |
| --- | --- | --- | --- |
| **Attributes** | | | |

| | | |
|---|---|---|
| **Gender** | | |
| Male | 42.8% | 48.8% |
| Female | 56.6% | 51.2% |
| | | |
| **Age** | | |
| 20-29 | 15.4% | 18.0% |
| 30-39 | 28.5% | 18.7% |
| 40-49 | 22.1% | 17.0% |
| 50-64 | 26.4% | 25.8% |
| 65+ | 7.6% | 20.5% |
| | | |
| **Household Size** | | |
| 1 | 13.1% | 25.0% |
| 2 | 29.3% | 28.5% |
| 3 | 20.3% | 17.3% |
| 4 | 22.4% | 17.1% |
| 5+ | 14.8% | 12.2% |
| | | |
| **Employment Type** | | |
| Employed full-time (30 hours or more per week) | 57.2% | 34.3% |
| Employed part-time (less than 30 hours per week) | 16.9% | 28.6% |
| Not employed | 25.9% | 37.2% |
| | | |
| **Workplace Arrangement** | | |
| Usual place of work | 43.7% | - |
| Worked at home & hybrid | 49.8% | - |
| No fixed workplace address | 6.6% | - |
| | | |
| **Household Income** | | |
| below $39,999 | 15.5% | 13.7% |
| $40,000 - $79,999 | 28.9% | 26.7% |
| $80,000 - $124,999 | 29.6% | 22.9% |
| $125,000 - $199,999 | 17.9% | 21.5% |
| $200,000 and above | 4.5% | 15.2% |
| Decline to answer/don't know | 3.4% | - |

## 4.1 Descriptive Statistics of Travel Behaviour from GTTS

The GTTS successfully captured the weeklong travel diary of 290 survey respondents. **Figure 3** presents the workflow of converting uploaded KML files into useful travel diaries. The GTTS captured a total of 13946 different GLH events. Of these events, 7560 were activities, and 6386 were trip legs. Each activity event had a location category annotated by GLH and an activity purpose validated by the respondent. Meanwhile, each trip leg had a mode inferred by GLH and a mode choice stated by the respondent.

Several rules were used to aggregate the trip legs into trips that recognize multimodal travel behaviour. The standalone legs not preceded or followed by another trip leg were separated and counted as single-mode trips with walking as access and egress mode. Then, sequential legs without activities in between were grouped, and each group's modes were analyzed. Groups of legs with only one type of travel mode or groups that only contained walking and one other travel mode were also described as single-mode trips. The rest of the trip leg groups contained several modes were then described as multimodal trips. The dataset contained 5,498 individual trips, with 5,441 single-mode trips using walking as their access and egress modes and 57 multimodal trips. The average trip rate was 2.71 per day. The following section will present a snapshot of travel behaviours revealed from the GTTS.

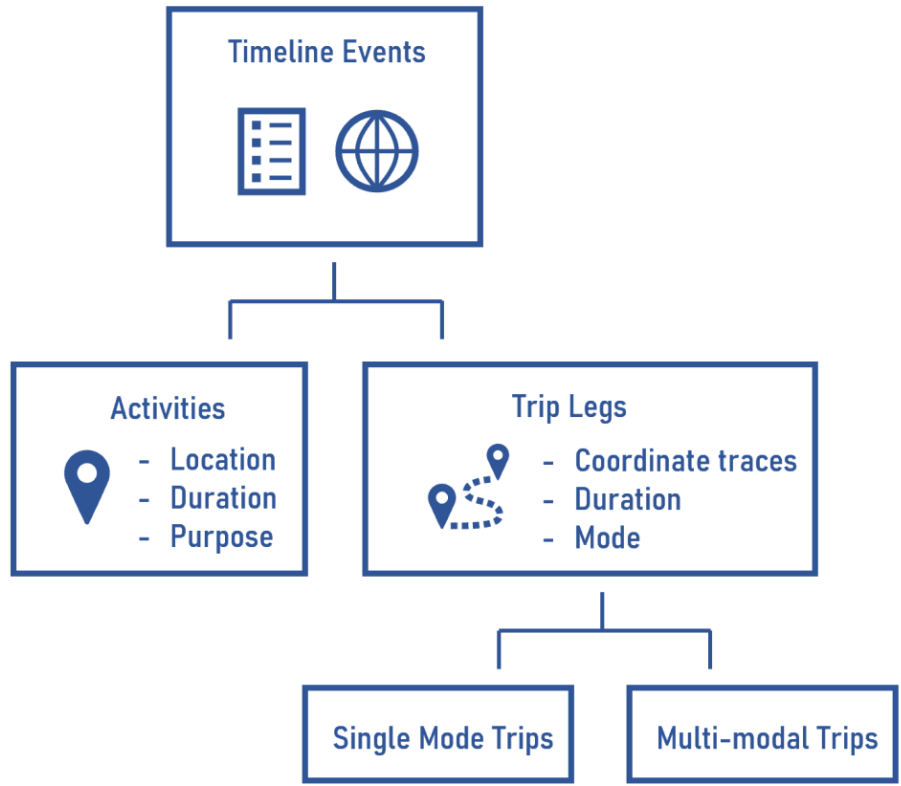

Figure 3. Dataset processing flow to convert Google Location History (GLH) into travel diaries

### 4.1.1 Trip Distances and Durations

Regarding trip distance, the GTTS dataset contained a wide distribution of trip lengths, as shown in **Figure 4**. The trip length distribution revealed a fair number of shorter trips under one kilometer. 8.5% of trips were under 500 meters. This highlighted that GLH could capture short-distance trips commonly underrepresented in CAWI or CATI-based travel surveys. **Figure 4** also shows that survey respondents' most common trip duration was between 10 and 30 minutes. Of the 5498 total trips recorded, 2122 (38.6%) had a duration between 10 and 30 minutes. The median duration was 15 minutes, and the mean was 52.3 minutes. This reflects that the long tail in the trip duration distribution heavily affect the mean of trip duration. The trip duration distribution in the GTTS were within similar range of durations from several other GPS-based travel diary studies. Kelly et al. (2013) analyzed twelve studies comparing travel information datasets from GPS versus self-reported diaries. They calculated that the average trip duration aggregated from the GPS datasets was 15.3 minutes compared to 19.7 minutes for self-reported travel diary data. In addition, in a study for the Oregon Household Travel Survey, Bricka et al. (2009) observed that the average trip duration amongst GPS-collected travel survey data was 13 minutes compared to 18 minutes using a traditional CATI-based household travel survey approach.

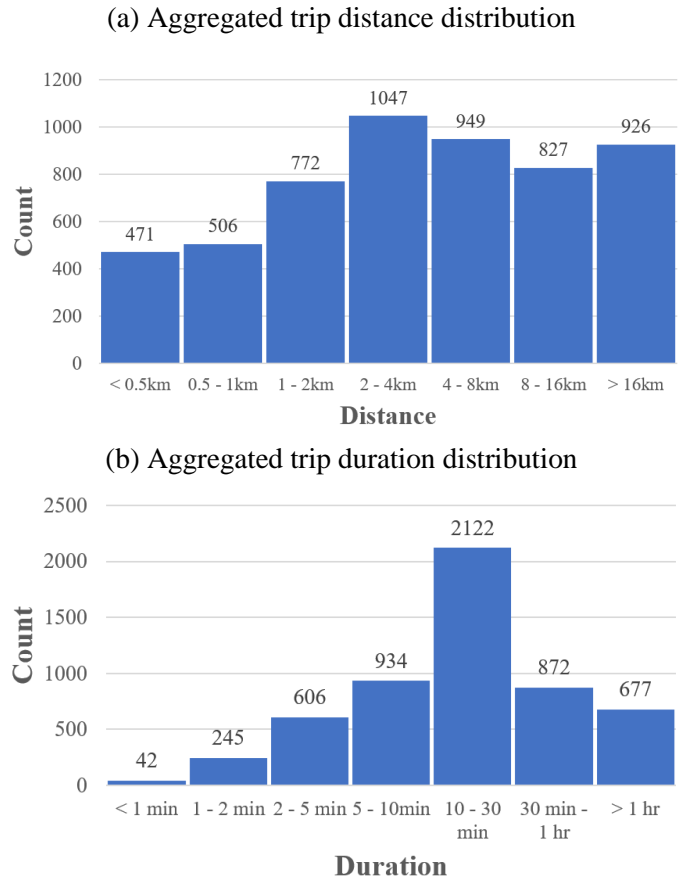

Figure 4. Aggregated trip distance and duration distribution

**Figure 5** presents the distribution of trip distances and durations by self-reported travel modes. The mode share was 71.1%, 8.2%, 17.7%, and 3% for automobile, transit, walking, and cycling, respectively. It can be observed that travellers primarily used automobiles and transit for long-distance travel. Meanwhile, active modes were primarily used for short trips. Most automobile trips were found in the range above 1 kilometer. Automobiles were favored for longest trips, with the most populous category being greater than 16 kilometers. Transit had a peak number of trips in the 4-to-8-kilometer range, with a sizeable amount greater than 8 kilometers. Cycle trips were most numerous in the 2-to-4-kilometer range, with more observations of short trips than trips longer than 4 kilometers. Finally, walking trips were the shortest, with more than 80% falling under 2 kilometers in length.

For the trip duration by each travel mode, most trips fell into the 10-to-30-minute duration category. It can be observed that GTTS respondents chose their mode of travel based on the trip's distance and the mode's speed. It appears that respondents attempted to keep their trips within 10 to 30 minutes in duration by choosing the appropriate mode. This is seen as the distributions of trip durations for automobiles, walking and cycling are very similar. On the other hand, most transit trips had at least over 30 minutes in duration. This can be attributed to transit system design and the relative inflexibility (e.g., fixed routes, limited transfer points, etc.) while riding the transit compared to other modes.

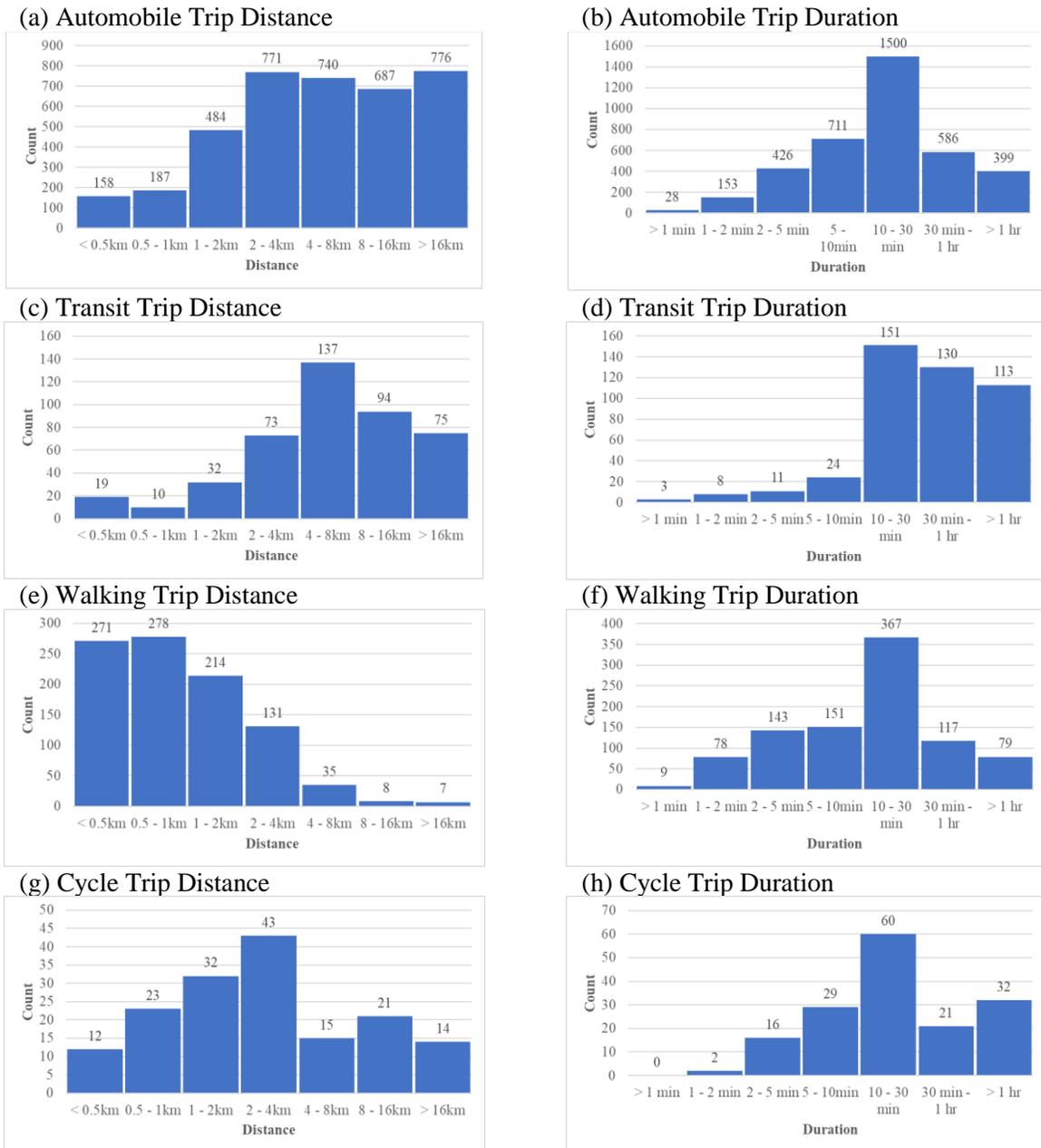

Figure 5. Trip Distributions by Modes

### 4.1.2 Activity Types & Durations

The GTTS captured 7560 activities. Among them, 3434 were home activities. These home activities included home working episodes. In total, 488 working-from-home activities were reported in the GTTS, resolving to 2.27 work-from-home activities per worker per week. It highlights the structural change in workplace arrangement in the post-pandemic era and the importance of explicitly collecting home working information in post-pandemic travel surveys (Wang et al., 2022; Wang et al., 2022b). **Figure 6** presents the composition of out-of-home activity types in the GTTS. Of the 4126 non-home activities, 26.0% were for shopping and errands, 19.3% were out-of-home work activities, 3% were for school, and the remaining 52% were other activities which included restaurants and services, meal pickup, day and overnight visits to friends and family, etc. It can be observed that individuals travelled for diverse purposes other than working, although 74.1% of the GTTS respondents were employed. This again supports the argument by Wang et al. (2022, 2022b) that flexible workplace arrangements dimmish the skeletal role of commuting in activity-based travel demand modelling approaches. Therefore, activity-based models should be adjusted for travel demand forecasting exercises in the post-pandemic era. Moreover, data collection efforts should explicitly collect information about working episodes at home.

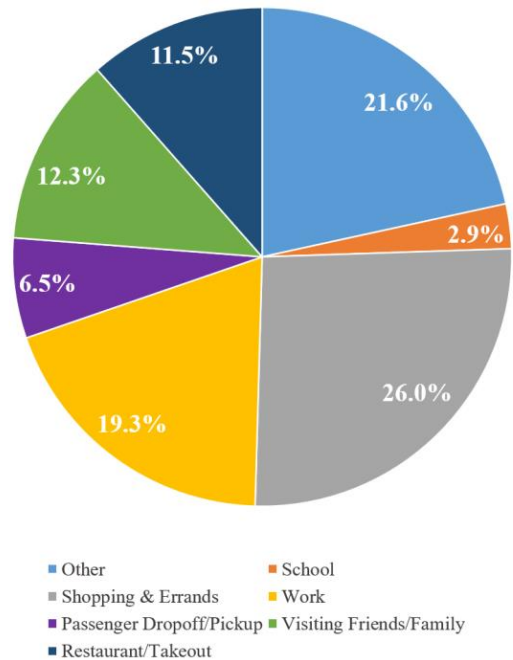

Figure 6. Out-of-home Activity Category Distribution

**Figure 7** presents distributions of the duration of work, school, shopping and errands and other activities. Most of the longer activities were for work and school, which were the primary activities that respondents should be engaged on the daily basis. Most work and school activities were over 4 hours in length. On the other hand, shopping trips and other discretionary activities had much less duration, distributed around average duration between 10 and 30 minutes. This is consistent with Cools et al. (2021). They found that GLH forgoes most activities that are less than 5 minutes. Those activities were trivial on individuals' timeline (e.g., go across street to grab a coffee during work). Retaining those activities will contribute to a much noisier dataset for activity-travel schedule analysis purposes. The review of activity duration distributions suggests that GLH could provide less noisy dataset containing non-trivial activity scheduling episodes on individuals' timeline.

a) Work Activity Duration

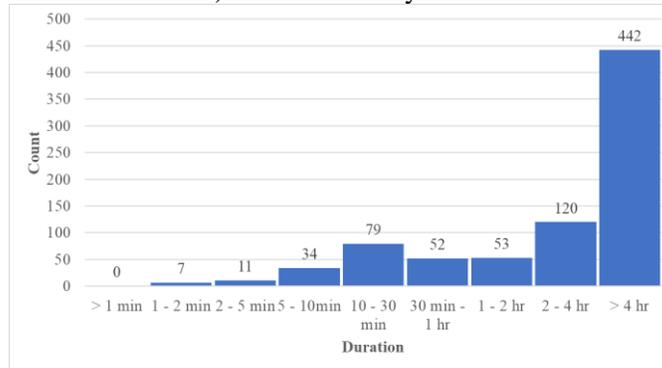

b) School Activity Duration

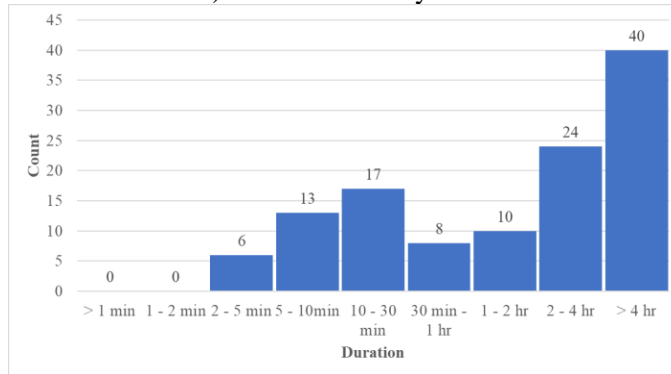

c) Shopping & Errands Activity Duration

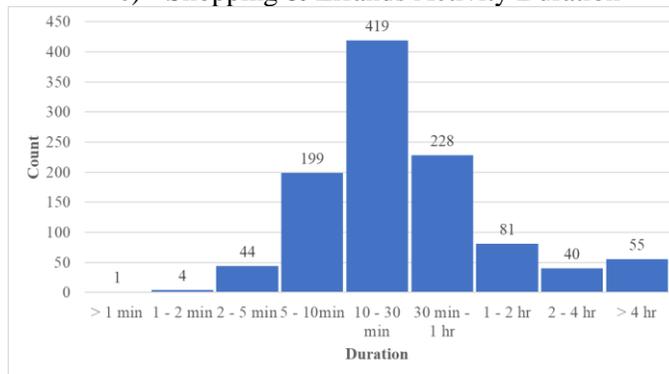

d) Other Activity Duration

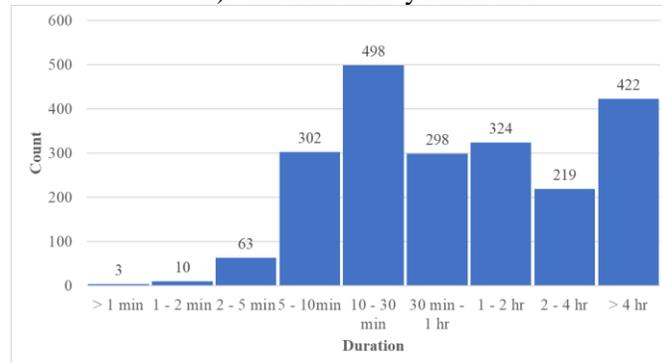

Figure 7. Activity Duration Distribution by Types

## 4.2 Validating GLH Mode Inference Model

Travel mode inference augments the GPS-based travel datasets (Shen & Stopher, 2014; Liu et al., 2021). The Google Location History (GLH) also annotated each trip leg with inferred travel mode. Meanwhile, the GTTS also asked respondents to validate their travel mode for each trip leg. This provides an invaluable opportunity to reference the accuracy of the preoperatory trip mode inferences algorithm developed by GLH. First, a confusion matrix is generated to examine the GLH inferred mode of each trip compared against the actual mode choice as validated by each respondent. Then, a binary logit model is developed to investigate influential factors of erroneous results inferred by the GLH.

**Table 2** reports the confusion matrix. For each mode type, a recall value is calculated, which considers the observations of the mode correctly predicted by Google divided by all observations of the mode validated by respondents. Similarly, the precision value is calculated by dividing the observations of a specific mode correctly predicted by GLH by all observations of the mode inferred by GLH. More specifically, the calculation of precision and recall follows:

$$Precision = \frac{trip\ legs\ correctly\ predicted\ as\ mode\ N}{total\ trip\ legs\ predicted\ as\ mode\ N} \quad (1)$$

$$Recall = \frac{trip\ legs\ correctly\ predicted\ as\ mode\ N}{all\ trip\ legs\ from\ mode\ N} \quad (2)$$

As inferred by GLH, the precisions for each mode were high except for motorcycles, which only have limited observation. This means that once GLH recognizes a specific travel mode, there is high confidence that the prediction is correct. To note, the two modes with the highest number of observations, automobile and walk, had precisions of 96.5% and 92.7%, respectively. Local, regional transit and cycle also had high precision values of over 80%. Surprisingly, taxi and ride-hailing trips had a precision of 100%.

On the other hand, the recall for automobiles, location transit, and active modes were all notably high. This means GLH could detect the categorical difference between the travel modes above. However, recall for regional transit, taxi, ride-hailing, and motorcycle was low. Regional transit had a lower recall value of around 46.9%. Taxi/ride-hailing and motorcycles have very low values of 9.4% and 12.5%. It can be postulated that GLH's algorithms may have difficulty separating trips of regional transit from trips utilizing local transit. This is backed by the fact that most of GLH's incorrect regional transit inferences were attributed to the local transit mode. For taxis/ride-hailing and motorcycles, it cannot be easy to separate these modes from automobiles as they all operate at similar speeds and rights of way in the road network. This fact, combined with the low number of observations, can cause the apparent lower capability of the algorithms in detecting these modes.

With insights from the confusion matrix, a binary logit model is developed to investigate exogenous factors associated with erroneous results inferred by the GLH mode inference algorithm. The independent variable in the binary logit model is the probability of GLH annotating a travel mode inconsistent with a self-validated mode. The inputs into the model are various trip characteristics and socioeconomic variables of the trip markers. Moreover, the self-reported travel mode is included in the final specification as dummy variables. This will handle the mode-specific effect on the probability of predicting incorrect results, leaving effects of the trip characteristics and socioeconomic variables unbiased to any specific mode. The pseudo-rho-square of the model is 0.43, highlighting good levels of model fit. The final model is summarized in **Table 3**. The dummy variable for automobile, transit, and cycle/walk had strong negative coefficients. This indicates that GLH is unlikely to estimate these modes inaccurately. The taxi/ride-hailing and motorcycle coefficients are kept to zero because they are not statistically significant while searching for the final specification.

Table 2. Google mode inference confusion matrix

| | | Google Inference | | | | | | | Recall (%) |
|---|---|---|---|---|---|---|---|---|---|
| | | Automobile | Local Transit | Regional Transit | Taxi/ Ridehail | Motorcycle | Cycle | Walk | |
| Respondent Validation | Automobile | 3642 | 30 | | | 1 | 10 | 45 | 97.7 |
| | Local Transit | 32 | 483 | 2 | | | 2 | 18 | 89.9 |
| | Regional Transit | 8 | 24 | 30 | | | | 2 | 46.9 |
| | Taxi/ Ridehail | 66 | 4 | 2 | 8 | | 1 | 4 | 9.4 |
| | Motorcycle | 4 | 1 | | | 1 | 1 | 1 | 12.5 |
| | Cycle | 7 | 5 | | | | 135 | 16 | 82.8 |
| | Walk | 16 | 4 | | | | 6 | 1095 | 97.7 |
| | Precision (%) | 96.5 | 87.7 | 88.2 | 100.0 | 50.0 | 87.1 | 92.7 | |

Regarding the trip characteristic variables, GLH mode inference is less accurate as average speed decreases. The algorithm is most inaccurate when the average speed is less than 5 km per hour. Conversely, longer trips are more likely to be correctly predicted by GLH. It is indicated by the positive sign of the dummy variable indicating trips that have lengths over 5 km.

Like trip characteristics, land-use characteristics also indicate that trips started from zones with larger population density have a negative effect on mode inference accuracy. A possible cause for this is that populated urban areas have more travel mode choices available. The higher popularity of transit trips in urban cores brought more uncertain for mode inference algorithms due to coinciding characteristics with other modes, as discussed by Liu et al. (2021). Moreover, congestion and other effects may cause each mode's speeds and movement characteristics to be similar. Stopher et al. (2008) noted difficulties distinguishing bus trips from car trips under congested conditions.

Regarding trip-makers characteristics, GLH appears less accurate with younger respondents under 30 years old. This may point to younger individuals having more complicated and emerging travel behaviours, such as using ride-hailing services which have the lowest recall value. Young & Farber (2019) found that younger trip-makers between 20 to 39 years old were primary users of ride-hailing services, with only 2% of ride-hailing users aged 60 and over. Lastly, full-time workers are more likely to be inferred with the correct travel mode by GLH. Workers might have more restrictive and constrained schedules emphasizing reliability and predictability in their mode choices. Thus, the mode choice pattern for workers might simplify the guesswork required by GLH.

Table 3. Binary logit model of the probability of Google Location History inferred travel modes different from self-reported modes in the GTTS

| | coefficient | t-statistics |
|---|---|---|
| ***Constant*** | 3.15 | 5.48 |
| ***Self-reported mode (dummy)*** | | |
| automobile | -7.01 | -12.97 |
| local transit | -6.17 | -11.36 |
| regional transit | -2.63 | -4.43 |
| taxi/ride-hailing | - | - |
| motorcycle | - | - |
| cycle & walk | -7.73 | -14.20 |
| | | |
| *Trip characteristics* | | |
| average speed (dummy) | | |
| < 5 km/hr. | 1.50 | 5.45 |
| 5 - 19 km/hr. | 0.93 | 4.31 |
| >= 20 km/hr. | - | - |
| | | |
| average distance (dummy) | | |
| >= 5 km | -0.70 | -3.15 |
| | | |
| *Land-use characteristics at the trip origin* | | |
| the logarithm of population density (thousand people per square km) | 0.16 | 2.03 |
| | | |
| *Trip-marker characteristics at the trip origin* | | |
| age < 30 (dummy) | 0.94 | 5.20 |
| full-time worker (dummy) | -0.45 | -2.50 |
| | | |
| Log-likelihood of the full model | -689.09 | |
| Log-likelihood of the constant-only model | -1209.84 | |
| Rho-square value against the constant-only model | 0.43 | |

## 5.0 Conclusion & future research

This study presents a travel survey method to collect and convert Google Location History (GLH) data into travel diaries suitable for transportation demand analysis. The proposed method does not require a proprietary GPS tracking application which is a barrier for researchers to collect smartphone-based GPS data. The potential reach of Google Location History based travel surveys is considerable as smartphones are extremely widespread, and the majority come with Google services installed. The practicality of the data collection method is demonstrated with the Google Timeline Travel Survey (GTTS) conducted in the Greater Toronto Area, Canada. The dataset obtained through the GTTS demonstrates great demographic representatives and provides detailed and accurate travel behavioural insights. The study also found that GLH annotated travel mode had high precision and recall accuracy for trips travelled by automobile, cycle and walk. The algorithm from GLH shows higher accuracy for trip speeds over 20 km per hour, larger than 5 km, and originated in less populated areas.

The proposed data collection method should be further developed to improve user-friendliness and response rate. It was recognized that more than 60% of the GTTS respondents completed the initial timeline setup during phase one but did not complete phase two. This may be due to barriers in comprehending the process of obtainment and submission of their GLH files. To improve, the survey process in the file uploading stage may be streamlined by providing clearer, interactive, or recorded instructions. Improvements to the survey methodology and interfaces would likely increase the response rate, providing a greater sample size for future surveys.

Overall, GPS-based travel surveys all suffered from low response rates. For example, Lynch et al. (2019) operated a travel survey with smartphone-based data collection in Minnesota with an overall rate of 2.0%. Calastri et al. (2020) also reported a similar response rate of 2% in a travel survey conducted in the UK that required a two-week smartphone tracking portion. As a result, there is rarely any GPS-based travel study collected more than four hundred samples. The only exception is the work of Calastri et al. (2020). The sample size of four hundred is the minimum requirement to yield reliable statistical representation for a study area with a population of over one million (Statistics Canada, 2010). Thus, developing survey methods to increase response and respondent retention rates for GPS-based travel surveys should be a collective objective for the transportation research and planning community.

**Acknowledgment**

The study was funded by an NSERC Discovery Grant. The authors are responsible for all results, interpretations, and comments on the paper.

**Author Contribution Statement**

The authors confirm their contribution to the paper as follows: Study conception and design: K. Wang, M. Li, K.M.N. Habib; Data collection: K. Wang, M. Li; Analysis and interpretation of results: K. Wang, M. Li, Y. Liu; Draft manuscript preparation: M. Li, K. Wang, Y. Liu; Overall project supervision: K.M.N. Habib. All authors reviewed the results and approved the final version of the manuscript.

Stopher, P. R. (1992). Use of an activity-based diary to collect household travel data. *Transportation*, *19*(2), 159–176. https://doi.org/10.1007/bf02132836

Strauts, E. (2010). Prediction of Cell Phone versus Landline Use in the General Social Survey. *Res Publica - Journal of Undergraduate Research*, *15*(1), 7. https://digitalcommons.iwu.edu/cgi/viewcontent.cgi?article=1152&context=respublica

Statistics Canada. (2023). *Profile table, Census Profile, 2021 Census of Population - Toronto [Census metropolitan area], Ontario*. https://www12.statcan.gc.ca/census-recensement/2021/dp-pd/prof/details/page.cfm?Lang=E&SearchText=toronto&GENDERlist=1,2,3&STATISTIClist=1&DGUIDlist=2021S0503535&HEADERlist=0

Statistics Canada. (2010). *Survey Methods and Practices* [Digital]. https://www150.statcan.gc.ca/n1/pub/12-587-x/12-587-x2003001-eng.pdf (Original work published 2003)

Wang, K., Liu, Y., Hossain, S., & Habib, K.M.N. (2023). *Who drops off web-based travel surveys? Investigating the implications of respondents dropping out of travel diaries during online travel surveys*. https://doi.org/10.21203/rs.3.rs-2512597/v1

Wang, K., Hossain, S., & Habib, K.M.N. (2022). What happens when post-secondary programmes go virtual for COVID-19? Effects of forced telecommuting on travel demand of post-secondary students during the pandemic. Transportation Research Part A: Policy and Practice, 166, 62–85. https://doi.org/10.1016/j.tra.2022.10.004

Wang, K., Mashrur, S. M., & Nurul Habib, K.M.N. (2022b). Developing a Flexible Activity Scheduling Model to InvestigatePost-Pandemic' Work Arrangement Choice" Induced Daily Activity-Travel Demands. SSRN Electronic Journal. https://doi.org/10.2139/ssrn.4078721

Wolf, J. (2006). Applications of new technologies in travel surveys. In Emerald Group Publishing Limited eBooks (pp. 531–544). https://doi.org/10.1108/9780080464015-029

Young, M., & Farber, S. (2019). The who, why, and when of Uber and other ride-hailing trips: An examination of a large sample household travel survey. Transportation Research Part A: Policy and Practice, 119, 383–392. https://doi.org/10.1016/j.tra.2018.11.018